\documentclass[prl,twocolumn]{revtex4}

\usepackage{graphicx}

\begin{document}

\title{Excited-state spectroscopy on a nearly-closed quantum dot via charge detection}

\author{J.M. Elzerman, R. Hanson, L.H. Willems van Beveren, L.M.K. Vandersypen, and L.P. Kouwenhoven}
\affiliation{Department of NanoScience and ERATO Mesoscopic Correlation Project, Delft University of Technology, Lorentzweg 1, 2628 CJ Delft, The Netherlands}

\date{\today}

\def\be{\begin{equation}}
\def\ee{\end{equation}}
\newcommand{\ket}[1]{\mbox{$|#1\rangle$}}
\newcommand{\mypsfig}[2]{\psfig{file=#1,#2}}

\begin{abstract}
We demonstrate a novel method for measuring the discrete energy spectrum of a quantum dot connected very weakly to a single lead. A train of voltage pulses applied to a metal gate induces tunneling of electrons between the quantum dot and a reservoir. The effective tunnel rate depends on the number and nature of the energy levels in the dot made accessible by the pulse. Measurement of the charge dynamics thus reveals the energy spectrum of the dot, as demonstrated for a dot in the few-electron regime.
\end{abstract}

\maketitle

Few-electron quantum dots are considered as qubits for quantum circuits, where the quantum bit is stored in the spin or orbital state of an electron in a single or double dot. The elements in such a device must have functionalities as initialization, one- and two-qubit operations and read-out~\cite{loss}. For all these functions it is necessary to have precise knowledge of the qubit energy levels. Standard spectroscopy experiments involve electron transport through the quantum dot while varying both a gate voltage and the source-drain voltage~\cite{Leo_review}. This requires that the quantum dot be connected to two leads with a tunnel coupling large enough to obtain a measurable current.

Coupling to the leads unavoidably leads to decoherence of the qubit: if the electron on the dot tunnels out and is replaced by another electron (whether by first- or second-order tunneling), the quantum information is irretrievably lost. Therefore, to optimally store qubits in quantum dots, the coupling to the leads must be made as small as possible. Furthermore, real-time observation of electron tunneling, important for single-shot read-out of spin qubits via spin-to-charge conversion, also requires a small coupling of the dot to the leads. In this regime, current through the dot would be very hard or even impossible to measure. Therefore an alternative spectroscopic technique is needed, which does not rely on electron transport through the quantum dot.

Here we present spectroscopy measurements using charge detection. Our method resembles experiments on superconducting Cooper-pair boxes and semiconductor disks which have only one tunnel junction so that no net current can flow. Information on the energy spectrum can then be obtained by measuring the energy for adding an electron or Cooper-pair to the box, using a single-electron transistor (SET) operated as a charge detector~\cite{Lafarge,Ashoori,Lehnert}. We are interested in the excitation spectrum for a given number of electrons on the box, rather than the addition spectra. We use a quantum point contact (QPC) as an electrometer~\cite{Field} and excitation pulses with repetition rates comparable to the tunnel rates to the lead, to measure the discrete energy spectrum of a nearly-isolated 1- and 2-electron quantum dot.

\begin{figure}[t]
\includegraphics[width=3.4in]{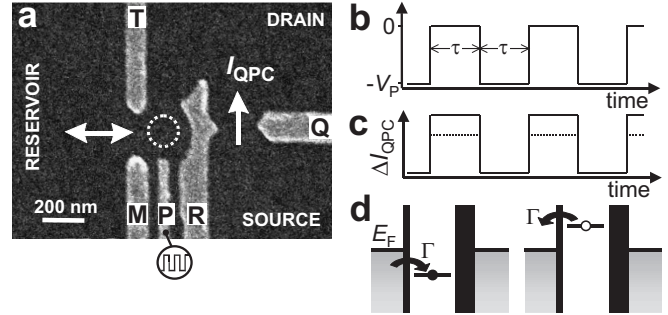}
\caption{(a) Scanning Electron Micrograph of a quantum dot and quantum point contact, showing only the gates used in the present experiment (the complete device is described in ref.~\protect\cite{Jeroen_ddot}). (b) Pulse train applied to gate $P$. (c) Schematic response in QPC current, $\Delta I_{QPC}$, when the charge on the dot is unchanged by the pulse (solid line) or increased by one electron charge during the ``high" phase of the pulse (dashed). (d) Schematic electrochemical potential diagrams during the high (left) and low (right) pulse phase, when the ground state is pulsed across the Fermi level in the reservoir, $E_F$.}
\label{fig1}
\end{figure}

The quantum dot and QPC are defined in the two-dimensional electron gas (2DEG) in a GaAs/Al$_{0.3}$Ga$_{0.7}$As heterostructure by dc voltages on gates $T, M, R$ and $Q$ (Fig.~\ref{fig1}a). The dot's plunger gate, $P$, is connected to a coaxial cable, to which we can apply fast voltage pulses. The QPC charge detector is operated at a conductance of about $e^2/h$ with source-drain voltage $V_{SD} = 0.2$ mV. All data are taken with a magnetic field $B_{\!/\!/} = 10$ T applied in the plane of the 2DEG, at an effective electron temperature of about 300 mK.

We first describe the procedure for setting the gate voltages such that tunneling in and out of the dot take place through one barrier only (i.e. the other is completely closed), and the remaining tunnel rate be well controlled. For gate voltages far away from a charge transition in the quantum dot, a pulse applied to gate $P$ (Fig.~\ref{fig1}b) modulates the QPC current via the cross-capacitance only (solid trace in Fig.~\ref{fig1}c). Near a charge transition, the dot can become occupied with an extra electron during the high phase of the pulse (Fig.~\ref{fig1}d). The extra electron on the dot reduces the current through the QPC. The QPC-response to the pulse is thus smaller when tunneling takes place (dotted trace in Fig.~\ref{fig1}c). We denote the amplitude of the difference between solid and dotted traces as the ``electron-response".

\begin{figure}[t]
\includegraphics[width=2.4in]{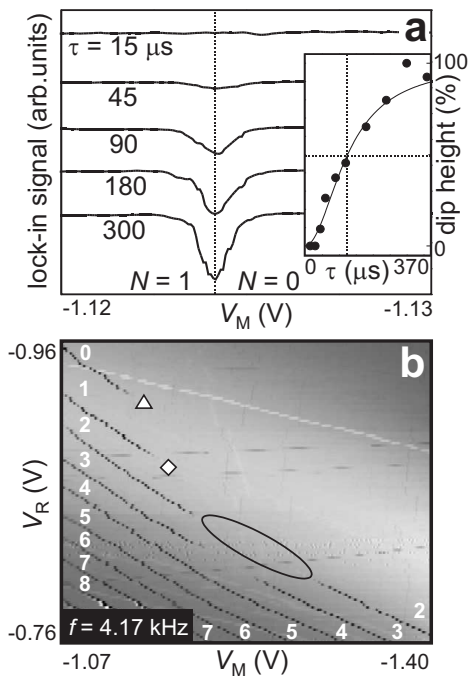}
\caption{Lock-in detection of electron tunneling. (a) Lock-in signal at $f = 1/(2\tau)$ versus $V_M$ for different pulse times, $\tau$, with $V_{P} = 1$ mV. The dip due to the electron-response disappears for shorter pulses. (Individual traces have been lined up horizontally to compensate for a fluctuating offset charge, and given a vertical offset for clarity.) (Inset) Height of the dip versus $\tau$, as a percentage of the maximum height (obtained at long $\tau$). Circles: experimental data. Dashed lines indicate the pulse time ($\approx120$ $\mu$s) for which the dip size is half its maximum value. Solid line: calculated dip height using $\Gamma = (40$ $\mu$s$)^{-1}$ (b) Lock-in signal in gray-scale versus $V_M$ and $V_R$ for $V_{P} = 1$ mV. Dark corresponds to dips as in (a), indicating that the electron number changes by one. White labels indicate the absolute number of electrons on the dot.}
\label{fig2}
\end{figure}

Now, even when tunneling is allowed energetically, the electron-response is only non-zero when an electron has sufficient time to actually tunnel into the dot during the pulse time, $\tau$. By measuring the electron-response as a function of $\tau$, we can extract the tunnel rate, $\Gamma$, as demonstrated in Fig.~\ref{fig2}a. We apply a pulse train to gate $P$ with equal up and down times, so the repetition rate is $f = 1/(2\tau)$ (Fig.~\ref{fig1}b). The QPC-response is measured using lock-in detection at frequency $f$, and is plotted versus the dc voltage on gate $M$. For long pulses (lowest curves) the traces show a dip, which is due to the electron-response when crossing the 0 to 1 electron transition. Here, $f \ll \Gamma$ and tunneling occurs quickly on the scale of the pulse duration. For shorter pulses the dip gradually disappears. We find analytically that the dip height is proportional to $1 - \pi^2/(\Gamma^2 \tau^2 + \pi^2)$, so the dip height equals half its maximum value when $\Gamma \tau = \pi$. This happens for $\tau \approx 120$ $\mu$s, so $\Gamma \approx(40$ $\mu$s$)^{-1}$. Using this value for $\Gamma$, we obtain the solid line in the inset to Fig.~\ref{fig2}a, which nicely matches the measured data points. 

We explore several charge transitions in Fig.~\ref{fig2}b, which shows the lock-in signal in gray scale for $\tau = 120$ $\mu$s, i.e. $f=4.17$ kHz. The slanted dark lines correspond to dips as in Fig.~\ref{fig2}a. From the absence of further charge transitions past the topmost dark line, we obtain the absolute electron number starting at 0. In the top-left region of Fig.~\ref{fig2}b, the right tunnel barrier, between gates $R$ and $T$, is much more opaque than the left tunnel barrier, between $M$ and $T$. Charge exchange occurs only to the left reservoir (indicated as "reservoir" in Fig.~\ref{fig1}a). Similarly, in the lower right region, charge is exchanged only with the drain reservoir. In the middle region, indicated for the 2 to 3 electron transition by an ellipse, both barriers are too opaque and no charge can flow into or out of the dot during the 120 $\mu$s pulse; consequently the electron-response becomes zero. By varying the voltages on gates $M$ and $R$, we can thus precisely set the tunnel rate through each barrier for each charge transition.

For spectroscopy measurements on a $N=1$ dot, we set the gate voltages near the 0 to 1 electron transition at the point indicated as $\bigtriangleup$ in Fig.~\ref{fig2}b. At this point, the dot is operated as a charge box, with all tunnel events occurring through just a single barrier. The pulse repetition rate is set to 385 Hz, so that the dip height is half its maximum value. The electron-response is then very sensitive to changes in the tunnel rate, which occur when an excited state becomes accessible for tunneling.

\begin{figure}[t]
\includegraphics[width=3.4in]{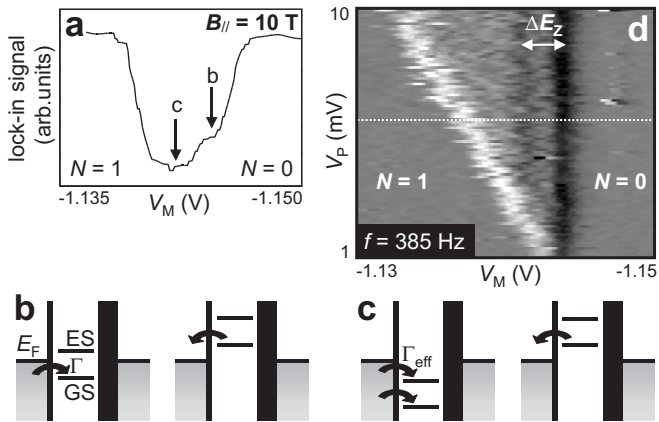}
\caption{Excited-state spectroscopy in a 1-electron dot. (a) Lock-in signal at $f = 385$ Hz versus $V_M$, with $V_{P} = 6$ mV. The dip is half the maximum value (obtained at low $f$ and small $V_{P}$) from which we conclude that $\Gamma \approx 2.4$ kHz. (b) Schematic electrochemical potential diagrams for the case that only the GS is pulsed across $E_F$. (c) Idem when both the GS and an ES are pulsed across $E_F$. (d) Derivative of the lock-in signal with respect to $V_M$ plotted as a function of $V_M$ and $V_P$ (individual traces have been lined up to compensate for a fluctuating offset charge). The curve in (a) is taken at the dotted line. The Zeeman energy splitting is indicated by $\Delta E_Z$.}
\label{fig3}
\end{figure}

Fig.~\ref{fig3}a shows the electron-response for a pulse amplitude larger than was used for the data in Fig.~\ref{fig2}. The dip now exhibits a shoulder on the right side (indicated by "b"), which we can understand as follows. Starting from the right ($N=0$), the dip develops as soon as the ground state (GS) is pulsed across the Fermi level $E_F$ and an electron can tunnel into the dot (Fig.~\ref{fig3}b). As $V_M$ is made less negative, we reach the point where both the GS and an excited state (ES) are pulsed across $E_F$ (Fig.~\ref{fig3}c). The effective rate for tunneling on the box is now the sum of the rate for tunneling in the GS and for tunneling in the ES, and as a result, the dip deepens (the electron-response increases). When $V_M$ is made even less negative, the one-electron GS lies below $E_F$ during both stages of the pulse, so there is always one electron on the dot. The electron-response is now zero and the dip ends.

The derivative of a set of curves as in Fig.~\ref{fig3}a is plotted in Fig.~\ref{fig3}d. Three lines are observed. The right vertical, dark line corresponds to the right flank of the dip in Fig.~\ref{fig3}a, the onset of tunneling to the GS. The slanted bright line corresponds to the left flank of the dip in Fig.~\ref{fig3}a (with opposite sign in the derivative) and reflects the pulse amplitude. The second, weaker, but clearly visible dark vertical line represents an ES. The distance between the two vertical lines is proportional to the energy difference between GS and first ES.

We identify the ground and excited state observed  in this spectroscopy experiment as the spin-up and spin-down state of a single electron on the quantum dot. For $B_{\!/\!/}=10$ T, the Zeeman energy is about 0.21 meV~\cite{Ronald_zeeman}, while the excitation energy of the first orbital ES is of order 1 meV. The distance between the two vertical lines can, in principle, be converted to energy and directly provide the spin excitation energy. However, it is difficult to determine independently the conversion factor between gate voltage and energy in this regime of a nearly closed quantum dot. Instead we take the measured Zeeman splitting from an earlier transport measurement~\cite{Ronald_zeeman} and deduce the conversion factor from gate voltage to energy, $\alpha = 105$ meV/V. This value will be used below, to convert the 2-electron data to energy.

\begin{figure}[t]
\includegraphics[width=3.4in]{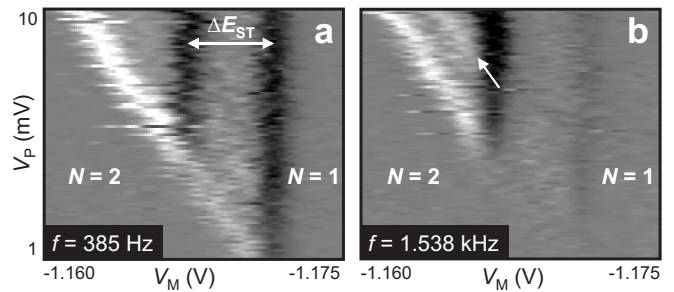}
\caption{Excited state spectroscopy in a 2-electron dot. (a) Similar to Fig.~\ref{fig3}d, but for the 1- to 2-electron transition. Again, $f = 385$ Hz.
We clearly observe the singlet-triplet splitting $\Delta E_{ST}$ (individual traces in (a) and (b) have been lined up). (b) Same experiment but with $f = 1.538$ kHz, which increases the contrast for excited states. An extra slanted line appears (arrow), corresponding to the $N = 1$ ES, spin-down.}
\label{fig4}
\end{figure}

Fig.~\ref{fig4}a shows pulse spectroscopy data for the $N = 1-2$ transition, taken with the gate settings indicated by $\diamond$ in Fig.~\ref{fig2}b. The rightmost vertical line corresponds to transitions between the $N = 1$ GS and the $N = 2$ GS (spin singlet) only. As $V_{P}$ is increased above 5 mV, the $N = 2$ ES (spin triplet) also becomes accessible, leading to an enhanced tunnel rate~\cite{tripletzeeman}. This gives rise to the left vertical line, and the distance between the two vertical lines corresponds to the singlet-triplet energy splitting $\Delta E_{ST}$. Converted to energy, we obtain $\Delta E_{ST} = 0.49$ meV.

Excitations of the $N = 1$ dot can be made visible at the $N = 1-2$ transition as well, by changing the pulse frequency to 1.538 kHz (Fig.~\ref{fig4}b). This is too fast for electrons to tunnel if only the GS is accessible, so the rightmost line almost vanishes. However, a second slanted line becomes visible (indicated by the arrow in Fig.~\ref{fig4}b), corresponding not to an increased tunnel rate into the dot (due to an $N = 2$ ES), but to an increased tunnel rate out of the dot (due to an $N = 1$ ES). Specifically, if the pulse amplitude is sufficiently large, either the spin-up or the spin-down electron can tunnel out of the 2-electron dot.

Similar experiments at the transition between 2 and 3 electrons, and for tunnel rates to the reservoir ranging from 12 Hz to 12 kHz, yield similar excitation spectra.

This work demonstrates that an electrometer such as a QPC can
reveal not only the charge state of a quantum dot, but also its
tunnel coupling to the outside world and the energy level spectrum
of its internal states. We can thus access all the relevant
properties of a quantum dot, even when it is almost completely
isolated from the leads.

We thank T. Fujisawa, S. Tarucha, T. Hayashi, T. Saku, Y. Hirayama and R.N. Schouten
for help and support. This work was supported by the DARPA-QUIST
program, the ONR and the EU-RTN network on spintronics.



\begin{thebibliography}{30}

\bibitem{loss}
D. Loss and D.P. DiVincenzo, Phys. Rev. A \textbf{57}, 120
(1998).

\bibitem{Leo_review}
L.P. Kouwenhoven, C.M. Marcus, P.L. McEuen, S. Tarucha, R.M. Westervelt, and N.S. Wingreen, in Mesoscopic Electron
Transport, v. 345 of NATO Advanced Study Institutes, Ser.
E: Applied Sciences,  L. L. Sohn, L. P. Kouwenhoven,
G. Sch\"on, Eds. (Kluwer Academic, Dordrecht, 1997).

\bibitem{Lafarge} P. Lafarge, H. Pothier, E.R. Williams, D. Esteve, C. Urbina, and M.H. Devoret, Zeitschrift f\"ur Physik B, {\bf 85}, 327 (1991).

\bibitem{Ashoori} R.C. Ashoori, H. L. Stormer, J. S. Weiner, L. N. Pfeiffer, S. J. Pearton, K. W. Baldwin, and K. W. West, Phys. Rev. Lett. \textbf{68}, 3088 (1992).

\bibitem{Lehnert} K.W. Lehnert, K. Bladh, L. F. Spietz, D. Gunnarsson, D. I. Schuster, P. Delsing, and R. J. Schoelkopf, Phys. Rev. Lett. \textbf{90}, 027002 (2003).

\bibitem{Field} M. Field, C. G. Smith, M. Pepper, D. A. Ritchie, J. E. F. Frost, G. A. C. Jones, and D. G. Hasko, Phys. Rev. Lett. \textbf{70}, 1311 (1993).

\bibitem{Ronald_zeeman} R. Hanson, B. Witkamp, L. M. K. Vandersypen, L. H. Willems van Beveren, J. M. Elzerman, and L. P. Kouwenhoven, Phys. Rev. Lett. \textbf{91}, 196802 (2003).

\bibitem{tripletzeeman} The expected Zeeman splitting of the triplet state is not resolved here.

\bibitem{Jeroen_ddot} J. M. Elzerman, R. Hanson, J. S. Greidanus, L. H. Willems van Beveren, S. De Franceschi, L. M. K. Vandersypen, S. Tarucha, and L. P. Kouwenhoven, Phys. Rev. B \textbf{67}, R161308 (2003).

\end{thebibliography}
\end{document}